\documentclass[aps,prb,twocolumn,groupedaddress,showpacs]{revtex4-1}

\usepackage{graphicx}
\usepackage{color}
\usepackage{ulem}
\usepackage{verbatim}
\usepackage{hyperref}

\begin{document}

\title{{\normalfont \textcolor{red}{\textsf{Version as published in Phys. Rev. B \textbf{102}, 176501 (2020), but now including an Appendix with a reply to the authors' published Reply}}}\\
\vspace{.5cm}Comment on ``Temperature range of superconducting fluctuations above $T_c$ in YBa$_2$Cu$_3$O$_{7-\delta}$ single crystals''}

\author{D. S\'o\~nora}
\author{J. Mosqueira}
\author{F. Vidal}

\affiliation{QMatterPhotonics Research Group, Departamento de F\'isica de Part\'iculas, Universidade de Santiago de Compostela, E-15782 Santiago de Compostela, Spain}

\date{\today}

\begin{abstract}
Contrary to the starting assumption of Grbi\'c et al. (Phys. Rev. B \textbf{83}, 144508 (2011)), here we will first argue that a 16~T magnetic field is not enough to quench all superconducting fluctuations above $T_c$ in YBa$_2$Cu$_3$O$_{7-\delta}$. We then conclude that through their measurements of microwave absorption these authors actually determine the ac-fluctuation magnetoconductivity at 16~T, instead of the zero-field ac paraconductivity as they pretend. So, the temperature proposed by Grbi\'c et al. for the onset of the superconducting fluctuations, $T'$, will correspond to the one at which the finite-field effects at 16~T become measurable in their experiments, and the actual fluctuations onset will be located well above $T'$. These conclusions, which also concern influential recent publications on that issue, are then confirmed by analyzing some of the Grbi\'c et al., data on the grounds of the Gaussian Ginzburg-Landau (GGL) approach for the finite-field (or Prange) fluctuation regime.
\end{abstract}

\maketitle

The starting assumption of the Grbi\'c and coworkers analysis of their interesting microwave absorption measurements in YBa$_2$Cu$_3$O$_{7-\delta}$ (YBCO) single crystals is \textit{that the field of 16 T is well sufficient to suppress all superconducting fluctuations above the zero field $T_c$}.\cite{Grbic11} Accordingly, these authors claim to extract the real part of the in-plane ac paraconductivity in zero magnetic field, as \textit{the difference of the curves} (of the conductivities) \textit{measured in zero field and in the field of 16 T}. The temperature at which this difference becomes measurable with their experimental resolution, denoted $T'$, is then identified with the onset of the superconducting fluctuations. To further support their conclusions, Grbi\'c and coworkers claim that the so-obtained zero field ac-paraconductivity may be quantitatively explained on the grounds of the GGL scenario.

In this Comment, we first stress that a magnetic field amplitude of 16 T is much smaller than different proposals for the upper critical magnetic field amplitude, $H_{c2}(0)$, of YBCO. As it is well known since the pioneering studies of M. Tinkham and coworkers in low temperature superconductors (LTSC),\cite{Gollub73,Tinkham} a field amplitude much smaller than $H_{c2}(0)$ is not enough to suppress the superconducting fluctuations  above $T_c$. As we will stress here, such a conclusion is general and will apply also to the high temperature superconductors (HTSC). As a consequence, instead of the real part of the zero field in-plane ac-paraconductivity, we will argue that the measurements of Grbi\'c and coworkers determine the real part of the total in-plane ac-magnetoconductivity at 16 T, and that the onset temperature of the superconducting fluctuations, $T_{onset}$, will be located well above $T'$. These conclusions are confirmed by analyzing, as an example, some of the data of Ref.~\onlinecite{Grbic11} for the overdoped sample on the grounds of the GGL approach for the finite-field (or Prange) fluctuation regime,\cite{Tinkham,Carballeira00,Rey19} and by taking also into account the frequency dependence of the ac conductivity.\cite{Tinkham,Peligrad03} Although Ref.~\onlinecite{Grbic11} has been published eight years ago, the interest and suitability of our present Comment is enhanced by the fact that some of the assumptions and/or conclusions of the paper that we now question are still being used in influential studies to support different phenomenological descriptions of the rounding effects around $T_c$,\cite{Bozovic18,Cyr18,Li13} which are in some cases contradictory.\cite{nota2}

The central aspect of the procedure used by Grbi\'c and coworkers was to approximate the zero-field in-plane paraconductivity, $\Delta\sigma_{ab}(\varepsilon,0)\equiv\sigma_{ab}(\varepsilon,0)-\sigma_{abB}(\varepsilon,0)$, by the in-plane magnetoconductivity, $\Delta\tilde\sigma_{ab}(\varepsilon,H)\equiv\sigma_{ab}(\varepsilon,0)-\sigma_{ab}(\varepsilon,H)$. In these expressions, $\sigma_{ab}(\varepsilon,H)$ and $\sigma_{abB}(\varepsilon,H)$ are respectively the as-measured and background (or normal-state) in-plane electrical conductivities, at a reduced temperature $\varepsilon\equiv\ln(T/T_c)$ and at a magnetic field, $H$, applied perpendicularly to the $ab$ planes.  However, this simple and well known approximation (see e.g., Refs.~\onlinecite{Tinkham,Rey19,Gap,Semba91,Ando02,Leridon07,Rullier11}, and references therein), which would allow to estimate the zero-field paraconductivity from two directly measurable observables, is applicable only if two conditions are fulfilled: First, if the normal state magnetoconductivity can be neglected, i.e., if $\sigma_{abB}(\varepsilon,0)\approx\sigma_{abB}(\varepsilon,H)$, and second, if it is used a field amplitude, $H_q$, large enough to quench the superconducting fluctuations at all reduced temperatures, i.e., $\sigma_{ab}(\varepsilon,H_q)=\sigma_{abB}(\varepsilon,H_q)$.      
													
The adequacy of the first approximation noted above, namely the smallness of the normal state magnetoconductivity when compared with the one associated with the superconducting fluctuations, has been earlier proved in different HTSC, including YBCO, at least at a qualitative level and up to moderate reduced temperatures and magnetic fields.\cite{Tinkham,Rey19,Gap,Semba91,Ando02,Leridon07,Rullier11} In what concerns the amplitude of $H_q$, the pioneering measurements of Gollub and coworkers of the fluctuation induced diamagnetism in several LTSC suggested that it is of the order of $H_{c2}(0)$, a result that these authors related to the shrinkage of the superconducting coherence length at high reduced magnetic fields and temperatures, already suggesting the generality of their finding.\cite{Gollub73,Tinkham} These conclusions were later confirmed by measurements in other LTSC \cite{Soto04} and also in a HTSC with a low $H_{c2}(0)$.\cite{Sugui12} The corresponding experimental results were accounted at a quantitative level on the grounds of the GGL scenario by introducing a so-called total-energy cutoff, which takes into account the limits imposed by the uncertainty principle to the shrinkage of the superconducting coherence length.\cite{Vidal02} Measurements of the dc paraconductivity under high magnetic fields (up to 60~T) in different HTSC, including YBCO compounds, also support $H_q\sim H_{c2}(0)$.\cite{Leridon07,Rullier11}

The question now is, therefore, if the $H_{c2}(0)$ values of the samples studied in Ref.~\onlinecite{Grbic11} are of the order or less than 16 T, the largest magnetic field used in these measurements. As already stressed in Tinkham's textbook,\cite{Tinkham} $H_{c2}(0)$ in HTSC \textit{is poorly defined, because of fluctuation rounding of the transition}. In fact, even in the case of the highly studied YBCO compounds around their optimal doping the discrepancies between different determinations of $H_{c2}(0)$ remain up to now very important, the proposals leading to field amplitudes between 100~T and 400~T.\cite{Tinkham,Rey19,Cyr18,Li13,Gap,Semba91,Ando02,Rullier11,Grissonnache14} For the most underdoped YBCO compound studied in Ref.~\onlinecite{Grbic11}, with $T_c = 57$ K, different proposals lead to $\mu_0H_{c2}(0)$ values between 30~T and 90~T, \cite{Ando02,Rullier11,Grissonnache14} in any case still much larger than the field amplitudes used in Ref.~\onlinecite{Grbic11}. For both compounds, the largest $H_{c2}(0)$ values are those extracted from the analysis of the superconducting fluctuations around $T_c$ on the grounds of the GL scenario, this last procedure probably being, as suggested by the above comment in Tinkham's textbook, the most adequate to determine $H_{c2}(0)$ in HTSC.

The qualitative analysis summarized above already suggest that $T'$ in Ref.~\onlinecite{Grbic11} actually corresponds to the temperature at which the finite-field effects at 16 T become measurable in their experiments. On the grounds of the GGL approach, these effects are expected to be appreciable when $\varepsilon$ becomes of the order of the reduced magnetic field $h\equiv H/H_{c2}(0)$,\cite{Gollub73,Tinkham,Carballeira00,Rey19,Gap} i.e.,
\begin{equation}
T'\approx T_c(1+h),
\end{equation}
an approximate relationship which will breakdown when $H\approx H_{c2}(0)$. In Ref. \onlinecite{Grbic11} the measured parameters are $T_c$ and $T'$(16 T), so the easiest way to check the applicability of Eq. (1) is to estimate the corresponding $H_{c2}(0)$ values. This leads to $\mu_0H_{c2}(0)\sim200$ T for sample OD89 (with $T_c=$89.4 K and $T'-T_c\approx 7$ K), and $\mu_0H_{c2}(0)\sim40$ T for sample UD57 (with $T_c=57.2$ K and $T'-T_c\approx 23$~K). These $H_{c2}(0)$ values are in reasonable agreement with the ones in the literature,\cite{Tinkham,Rey19,Gap,Semba91,Ando02,Rullier11,Grissonnache14} taking into account the error sources affecting both the experimental parameters and Eq.~(1), in particular the assumption that the normal-state magnetoconductivity may be neglected. Note also that the seemingly much wider temperature range of the superconducting fluctuations observed in the most underdoped sample with $T_c\approx 57$~K (when compared to the widths observed in the almost optimally doped samples), a result claimed by Grbi\'c and coworkers\cite{Grbic11} as the \textit{most intriguing  in the current controversy about the nature of the pseudogap in deeply underdoped HTSC}, may be easily explained by just taking into account in Eq.~(1) the much lower value of the upper critical field in this deeply underdoped sample. 

The finite field effects predicted by the GGL approach may also easily explain the central result above $T_c$ in Fig.~6 of Ref.~\onlinecite{Grbic11}: The overlapping a few degrees (7-8~K) above $T_c$ (denoted $T'$) of the two curves of the in-plane complex conductivity of sample UD87 measured under 0 and 16~T. This result just confirms the presence of the expected field effects below $T'(16$ T), and it does not exclude at all the presence of appreciable superconducting fluctuations in the entire temperature region above $T'$ covered by these data. So, the conclusion of Ref.~\onlinecite{Grbic11} that \textit{by taking the difference of the conductivities in the zero field and in 16~T field, one may constrict a reliable procedure for extracting the pure superconducting fluctuation contribution to the conductivity avove the zero field $T_c$}, is unfounded.

A quantitative check of the crude conclusions summarized above may be done on the grounds of the GGL approach by analyzing, as an example, the results on the real part (see Ref.~\onlinecite{notareal}) of the ac measurements presented in Fig.~9(a) of Ref.~\onlinecite{Grbic11} for the overdoped sample OD89, which is the closer to the prototypical optimally-doped YBCO. As commented above, these data actually correspond to the ac in-plane magnetoconductivity, $\Delta\tilde\sigma_{ab}(\varepsilon,\rm{16~T})$. By neglecting the normal state magnetoconductivity when compared with the one associated with the superconducting fluctuations, these data may be then approximated as,
\begin{eqnarray}
&&\sigma_{ab}(\varepsilon,0)-\sigma_{ab}(\varepsilon,16 T) \nonumber\\
&&\approx S(\omega,T)\left[\Delta\sigma_{ab}(\varepsilon,0)-\Delta\sigma_{ab}(\varepsilon,16T)\right].
\end{eqnarray}
For the pre-factor $S(\omega,T)$, which takes into account the high-frequency influence on the ac conductivity, we have used Eq.~(12) of Ref.~\onlinecite{Peligrad03}, as in Ref.~\onlinecite{Grbic11}. In turn, for the dc paraconductivity, $\Delta\sigma_{ab}(\varepsilon,H)$, we have used Eq.~(8) in Ref.~\onlinecite{Rey19}. 

The solid line in Fig. 1 corresponds to the best fit of Eq.~(2) to the $\sigma_{1ab}$ data in Fig. 9(a) of Ref.~\onlinecite{Grbic11}. In doing this fitting we have excluded the data within the transition width $\Delta T_c$ reported in Ref.~\onlinecite{Grbic11} ($\varepsilon<\Delta T_c/T_c$, dashed bar) probably affected by $T_c$ inhomogeneities\cite{nota}, and the only free parameter was the cutoff $\Lambda$ arising in $S(\omega,T)$. For the remaining parameters, those arising in $\Delta\sigma_{ab}$ (the in-plane and transverse coherence lengths, the relative GL relaxation time, and the total-energy cutoff), we have used the values recently obtained in optimally doped YBCO through measurements of the precursor diamagnetism and the dc paraconductivity and magnetoconductivity, summarized in Table 2 of Ref.~\onlinecite{Rey19}: $\xi_{ab}(0) = 1.1$ nm, $\xi_c(0) = 0.11$ nm, $\tau_{rel} =1$ and $\varepsilon_c=0.55$. Our procedure then assures a crucial check of consistency with previous dc studies.\cite{Ohashi09} 
Another important aspect of our present analysis is that it leads to $\Lambda=0.23$, in reasonable agreement with the theoretical expectations\cite{Peligrad03} (see also below). 
Note also that, as the $\varepsilon$-region where the field effects at 16 T become relevant is relatively close to $T_c$, the resulting fit will not be appreciably affected by a cutoff in the dc paraconductivity (although the total-energy cutoff is crucial to precisely locate $T_{onset}$ in the GGL scenario\cite{Rey19,Vidal02}). 
  
\begin{figure}[t]
\begin{center}
\includegraphics[scale=.55]{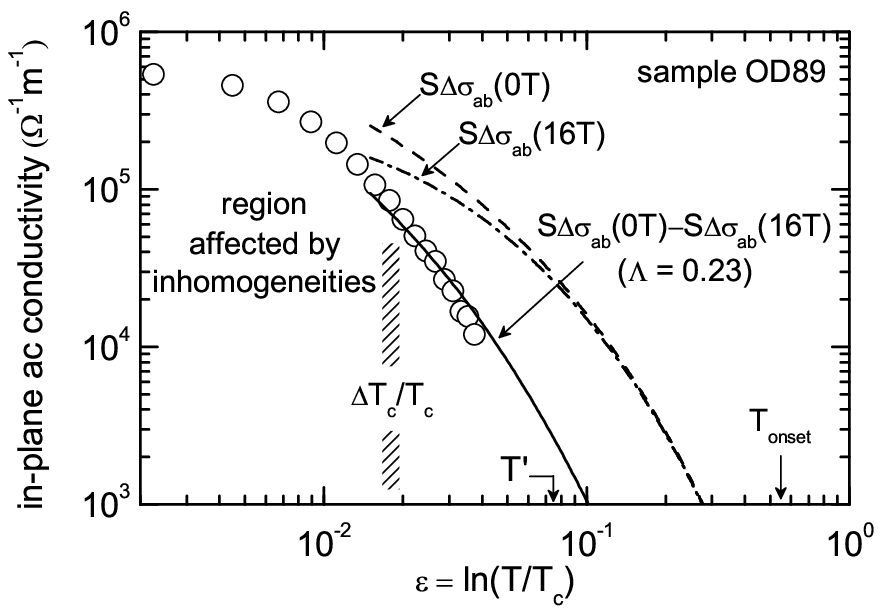}
\caption{The solid line corresponds to the best fit of the in-plane magnetoconductivity given by Eq.~(2) to the $\sigma_{1ab}$ data in Fig.~9(a) of Ref.~\onlinecite{Grbic11} for the overdoped sample OD89. This solid line was approximated as the difference between the in-plane ac paraconductivity measured without field and with 16 T (dashed lines). This fit excludes data within the transition width (below $\sim\Delta T_c/T_c$), which may be deeply affected by $T_c$-inhomogeneities. In this GGL scenario, the well-defined onset temperature of the superconducting fluctuations is noted $T_{onset}$ (which corresponds to a total-energy cutoff of 0.55), whereas $T'$ was proposed in Ref.~\onlinecite{Grbic11}. However, as it may be appreciated, this last corresponds instead, well within the experimental resolution, to the onset of the field effects at 16 T (i.e., when $\varepsilon\approx h$). For other details see the main text.}
\end{center}
\end{figure}

Taking into account the crude approximation used to estimate the high-frequency effects (although similar to the one used in Ref.~\onlinecite{Grbic11}), and that the only free parameter was the cutoff $\Lambda$ arising in $S(\omega,T)$, the results summarized in Fig.~1 may be considered as reasonable, and compare favorably with the analysis performed in Ref.~\onlinecite{Grbic11}. This conclusion is illustrated in Fig.~2, where the same data were compared in Ref.~\onlinecite{Grbic11} with the zero-field in-plane paraconductivity, with all fitting parameters free and without excluding the data-points closer to $T_c$. As it may be seen, the agreement is good in the region surely affected by $T_c$-inhomogeneities but worse at higher-$\varepsilon$.\cite{nota} Moreover, the fitting parameters present non reasonable values: $\Lambda=0.026$ is anomalously smaller than the unity, and the $\varepsilon$-onset of the critical fluctuation region ($\Gamma=0.21$) is one order of magnitude larger than in the literature.\cite{Gap,eLG}

It is finally worth noting that, the $T'$ value proposed in Ref.~\onlinecite{Grbic11} agrees, well within the experimental uncertainties, with the onset of the predicted field effects on the paraconductivity at 16~T (see Fig.~1). The implications of the precise location of $T_{onset}$, in particular in relation to the so-called pseudogap temperature, is still at present a debated central aspect of the phenomenological descriptions of the superconducting transition in HTSC (comments on this issue may be seen in Sections 5.2 and 5.3 of Ref. \onlinecite{Rey19}, and also in Refs. \onlinecite{Bozovic18,Cyr18} and references therein). 

\begin{figure}[t]
\begin{center}
\includegraphics[scale=.55]{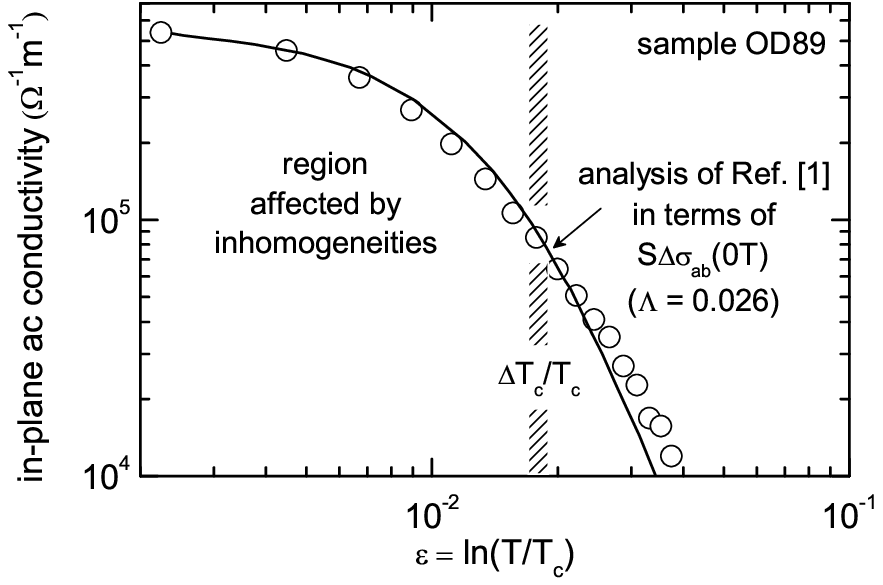}
\caption{Analysis performed in Ref.~\onlinecite{Grbic11} (see the Fig. 9(a) there) of the same data that we have analyzed in our Fig.~1. However, in this case these data were compared with the zero-field in-plane ac-paraconductivity, with all fitting parameters free and without excluding the data-points closer to $T_c$. One may appreciate here that the agreement is good in the region probably affected by $T_c$-inhomogeneities, but worse at higher $\varepsilon$ (see the main text and the note in Ref.~\onlinecite{nota}).}
\end{center}
\end{figure}

In conclusion, some of the interesting microwave absorption measurements of Ref. \onlinecite{Grbic11} in a overdoped YBCO sample have been explained quantitatively, and consistently with high quality dc measurements in similar samples, on the grounds of the GGL approach for conventional superconducting fluctuations. In particular, our results confirm that the reduced-temperature range of the superconducting fluctuations above $T_c$ extends well beyond the one proposed in Ref.~\onlinecite{Grbic11}, almost one order of magnitude in the case of the overdoped YBCO sample. These conclusions also concern several of the proposals, in some cases contradictory, of Refs.~\onlinecite{Bozovic18,Cyr18,Li13}, and enhance the interest of extending to other HTSC and to other observables, quantitative analysis on the grounds of the GGL scenario. 

\textit{Note added in proof}. Two other references are relevant to this discussion [\onlinecite{Pelc20,Maza91}].

This work was supported by the Agencia Estatal de Investigaci\'on (AEI) and Fondo Europeo de Desarrollo Regional (FEDER) through projects FIS2016-79109-P and PID2019-104296GB-I00, and by Xunta de Galicia (grant GRC no. ED431C 2018/11).

\newpage

\textbf{APPENDIX:  Reply to the author's Reply}
\vspace{.5cm}

In their original paper, Grbi\'c et al. considered the onset temperature estimated in their microwave measurements, $T'$, as the onset temperature of ``\textit{all superconducting fluctuations above the zero field $T_c$}''. However, in the Reply to our Comment, a new physical meaning is attributed to $T'$, this time referring to the ``\textit{onset temperature of the superconducting fluctuations in the microwave spectral range above $T_c$}'', which ``\textit{do not preclude superconducting fluctuations with frequencies outside the microwave window at elevated temperatures}''. Additionally, the cited authors avoid any further comparison with theoretical models. The aim of this Appendix is to show that the reasoning in that Reply against our Comment remains unfounded, and that the new physical meaning assigned to $T'$ is still incorrect. Nevertheless, we must stress that when correctly analyzed, their experimental data agree with the DC measurements of the paraconductivity in optimally doped YBCO and are explained by the GGL scenario.

\renewcommand{\thesubsection}{A\arabic{subsection}}
\setcounter{figure}{0}
\subsection{On the quenching of the superconducting fluctuations by low reduced magnetic fields, and background estimations}

The only quantitative argument in the Reply proposed by Grbi\'c et al. is based on Fig.~1, where new data for a slightly overdoped sample are presented. They claim that for temperatures above 91~K, the difference between the two curves under magnetic fields of 12~T and 16~T ``\textit{falls below our experimental limit (0.1~ppm). Hence both high field data can be used as background for $T>91$~K}''. This argument is, in fact, the same as their original paper's. As it was already stressed in our Comment, the apparent coincidence of these curves at temperatures slightly above $T_c$ is because the finite field effects associated with relatively low reduced fields are present only near $T_c$. A rough approximation of these effects can be obtained using Eq.~(1) of our Comment: For the slightly overdoped YBCO sample $\mu_0H_{c2}(0)\sim300$~T, as it follows from a coherence length amplitude of $\xi(0)\sim1$~nm, and the onset of finite-field effects is about 92.5~K and 94~K under a 12~T and 16~T magnetic field, respectively. These values are in good agreement with the temperatures at which the 12~T and 16~T data series in Fig.~1 of the Reply merge with the 0 T measurement. 

It is also worth noting that, in the Reply, it is mentioned that ``\textit{the same criterion was used in the pioneering work of Tinkham and co-workers [4,5], evoked by the authors of the preceding Comment [1], for the determination of zero magnetization (background) in their experiments}''. But this is not true: In these works Tinkham and coworkers take advantage of the temperature independence of the magnetization in the normal state, and for each applied field use as background the constant magnetization value for $T \gg T_c$.

\renewcommand{\thefigure}{A\arabic{figure}}
\setcounter{figure}{0}

\begin{figure}[t]
\begin{center}
\includegraphics[scale=.5]{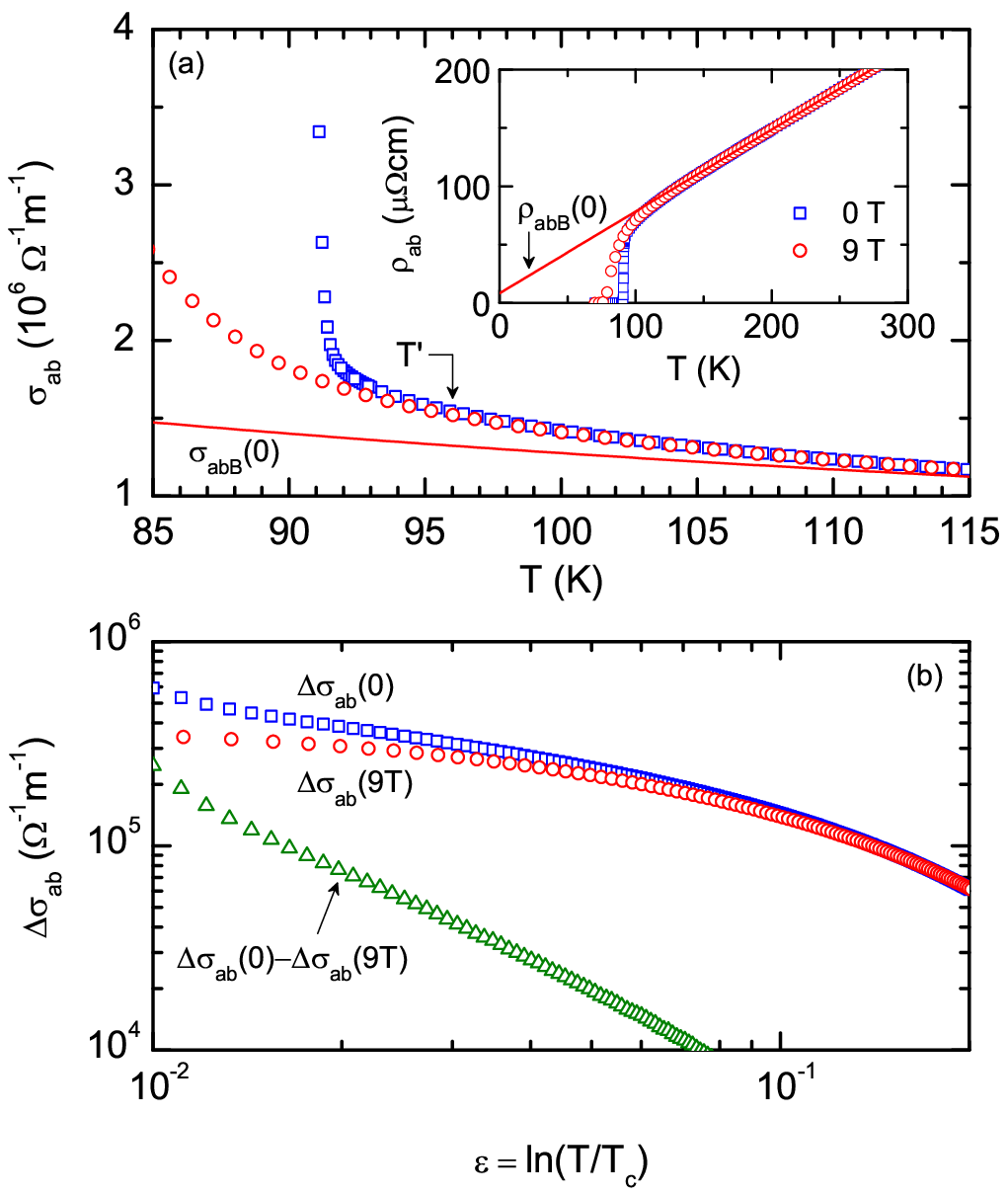}
\caption{(a) Temperature dependence of the as-measured in-plane DC conductivity under zero magnetic field (blue open squares) and under an external magnetic field of 9~T (red open circles), of a YBCO film, with $T_c = 91.1$~K. These data were taken from Fig.~2(a) of Ref.~5. The solid line is the background conductivity, obtained by extrapolating through the superconducting transition the normal state resistivity, as indicated in the inset, where the corresponding resistivity curves are also shown. 
Following Grbi\'c et al. criterion (still maintained in their Reply), the onset temperature for the superconducting fluctuations ($T'$ in the figure) would be in flagrant disagreement with the comparison between the measured and the background conductivities shown in this figure, and with the GGL approaches.
It is particularly enlightening the comparison of  these curves with those of Fig.~6 of the Grbi\'c and coworkers original article. (b) Log-log plot of the in-plane DC paraconductivity obtained from the data shown in (a), together with the actual in-plane DC fluctuation magnetoconductivity under a magnetic field of 9 T. For clarity, in some of these datasets only some of the data points are shown.}
\end{center}
\end{figure}

\subsection{Comparison with DC measurements}

Two curves of the in-plane DC conductivity of an optimally doped YBCO film ($T_c = 91.1$~K) are presented in Fig.~A1. These curves were extracted from Fig.~2 of Ref.~5 and correspond to DC measurements of the zero-field resistivity (blue open squares) and under a 9~T magnetic field (red open circles). The latter data were obtained at a higher magnetic field than the cited authors consider ``\textit{sufficient to suppress superconductivity at temperatures above the zero-field $T_{c0}$}''. From the resistivity data, shown in Fig.~A1(a) inset, the background resistivity (solid line) can be calculated as an extrapolation of the linear normal-state resistivity well above $T_c$ (see Ref.~5 for more details). As it can be seen in Fig.~A1(a), the linear representation shows a rather similar result to the one obtained by Grbi\'c et al. in their original paper (see Fig.~6 therein): at about 5~K above $T_c$, the two curves visually merge. However, considering that these measurements were taken in DC, and following their criteria, it would follow that $T'$ is the onset temperature of all (not just in the microwave spectral range) superconducting fluctuations above $T_c$, and the curve obtained for a 9~T field would correspond to the background conductivity. This is in stark contrast with the estimated background conductivity (solid line in Fig.~A1(a)).

The paraconductivities, obtained from the conductivity and background curves shown in Fig.~A1(a), are presented in Fig.~A1(b) in a log-log scale, where the fluctuation magnetoconductivity, obtained as the difference between the zero-field and the 9~T paraconductivity, is also shown (green triangles). Taking into account the amplitude reduction associated with the high-frequency cutoff effects arising in the microwave measurements, the comparison with the results in the original article by Grbi\'c et al. provides a further confirmation that ``\textit{... these authors actually determine the ac fluctuation magnetoconductivity at 16~T, instead the zero-field ac paraconductivity as they contend}'', as it was stressed in the abstract of our Comment.

\subsection{On the presence of $T_c$-inhomogeneities} 

Another major criticism that can be made is that in their Reply Grbi\'c et al. have still neglected the role played by the presence of $T_c$ inhomogeneities. This is perhaps surprising, because in a recent article (Ref.~24 in our Comment), co-authored by one of the authors of the commented paper, it is proposed that the resistivity rounding above $T_c$ is dominated by the presence of intrinsic inhomogeneities in these compounds. The intrinsic character of these inhomogeneities may be questionable, but their presence would support the disagreement with the GGL approach that we observe below $\varepsilon\approx0.02$, and are difficult to reconcile with the excellent agreement below that $\varepsilon$-value claimed by Grbi\'c and coworkers (see Figs.~1 and 2 of the Comment). Although it should be noted that $T_c$ inhomogeneities alone cannot account for the observed resistivity rounding at higher $\varepsilon$, at least in optimally doped YBCO, as it has been concluded a long time ago by using the effective-medium theory (see Ref.~25 in our Comment).


\begin{thebibliography}{100}
\bibitem{nota2}For instance, in spite that several of the articles in Refs.~\onlinecite{Bozovic18,Cyr18,Li13} use the hypothesis and/or conclusions of Ref.~\onlinecite{Grbic11} that we question here, their proposals to explain the observed effects above $T_c$ include physical mechanisms as different as conventional superconducting fluctuations, phase fluctuations, or percolation processes.
\bibitem{notareal}As it may be seen in Figs.~7 to 9 of Ref.~\onlinecite{Grbic11}, the imaginary part of the complex conductivity ($\sigma_2$) decays much faster with $\varepsilon$ than the real one ($\sigma_1$). This makes $\sigma_2$ more strongly affected by $T_c$ inhomogeneities, that are expected to extend up to $\varepsilon\approx\Delta T_c/T_c\approx0.02$ in sample OD89. Moreover, the negligibly small values of $\sigma_2$ at high $\varepsilon$ makes difficult to reliably detect the fluctuations onset. Finally, $\sigma_1$ may be more easily compared with the dc conductivity, a crucial check of consistency.
\end{thebibliography}
\end{document}